\documentclass[preprint,aps]{revtex4}

\usepackage{graphicx}
\usepackage{dcolumn}
\usepackage{bm}

\begin{document}


\title{Direct sampling of optical coherence using quantum interference}

\author{Jungsang~Kim and David~J.~Brady}
\affiliation{Fitzpatrick Institute for Photonics and Electrical \& Computer Engineering Department \\
Duke University, Box 90291, Durham, NC 27708}
\date{\today}

\begin{abstract}
We describe a detector that measures the mutual coherence of two optical fields directly using quantum interference, free from photon noise of the individual irradiances. Our approach utilizes Raman transition in an atomic system where the state evolution is driven by the mutual coherence of the fields interacting with the atoms.  Feedback control is used to balance the interaction of the fields being characterized, providing a measure of the mutual coherence. We show that the sensitivity of the coherence measurement can be enhanced significantly above that of conventional interferometric methods when the mutual coherence of the two fields is weak.

\end{abstract}

\maketitle

The assumption that detectors measure irradiance is a foundational principle of modern optical systems. System design and measurement statistics are radically different for irradiance detectors in comparison with the field detectors used at lower frequencies. For example, optical imaging systems use lenses as analog image formation devices whereas radar systems use phased arrays and digital post processing. Relative to wavelength scale, the radar devices are much thinner and the resulting images are of much higher quality.
Similar lensless computational imaging strategies were previously demonstrated at optical frequencies based on measuring the mutual coherence function across an aperture~\cite{MarksScience1999}, but two-point coherence measurement by classical interfereometry is corrupted by background terms that make this strategy useless for all but the simplest images~\cite{basty03}. In this paper, we propose a solution to this problem by revising the assumption that irradiance is the only direct optical observable.  We consider a quantum system whose evolution is driven by the mutual coherence of fields it interacts with, and propose a method to estimate the mutual coherence by measuring the optical control signal necessary to maintain the quantum system in a static state. This is analogous to measuring the radio frequency field in an antenna by characterizing the current necessary to maintain a boundary condition. We show that it is possible to measure the mutual coherence of two fields independent of the photon shot noise of each field.

The {\em mutual coherence} between two Maxwell fields $E_1$ and $E_2$ drawn from two space time points is defined as $
\Gamma_{12} (\tau) \equiv \langle E_1 (t) E^*_2 (t+\tau)\rangle /Z_0$, where $Z_0$ is the impedance of vacuum. The irradiance of the individual fields correspond to diagonal elements $I_i=\Gamma_{ii}$\cite{MandelWolf}. Like the Maxwell fields $\Gamma_{12}$ may be calculated anywhere in a source and occlusion free region from its value on an enclosing boundary. Unlike the Maxwell fields, $\Gamma$ may be completely specified from intereferometric irradiance measurements. 
The mutual coherence of two incoherent fields can be measured using classical optical interferometry~\cite{HillAO1997,ZmuidzinasAO2003}. Techniques that seek to characterize the coherent ${\bf E}$ field, such as holographic imaging in spatial domain~\cite{SalehTeich1991} or
optical homodyne and heterodyne measurements in time  domain~\cite{YuenOL1983} use irradiance detectors to characterize $\Gamma_{12}$.

\begin{figure}[tbp]
        \centering
                \includegraphics[width=8cm]{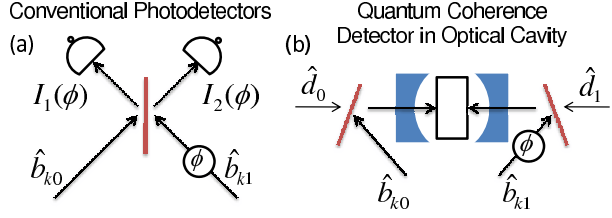}
                \caption{Schematic for sensing mutual coherence using (a) a beam splitter and classical irradiance detectors and (b) quantum coherence detector.} 
\label{InterferometerScheme}
\end{figure}

Classical interferometry for measuring $\Gamma_{12}$ is illustrated in Fig. \ref{InterferometerScheme}a. The two optical fields are described by a set of annihilation operators $\hat{b}_{k0}$ and $\hat{b}_{k1}$ with a quasi-monochromatic spectrum featuring a center frequency of $\omega_c$, bandwidth of $\Delta \omega$, and relative phase shift of $\phi$. We assume that the spectrum of the two fields is identical, so that $\Gamma_{11}=\Gamma_{22} \equiv I_0$ and $\langle \hat{b}_{ki}^\dagger\hat{b}_{k'i} \rangle= n_{k}\delta_{kk'}$, where $n_k$ is the mean photon number in the mode $k$. The normalized mutual coherence $\zeta \equiv \Gamma_{12} / I_0$ characterizes the level of coherence between the modes $\hat{b}_{k0}$ and $\hat{b}_{k1}$. The irradiance on the detectors are 
\begin{equation}
  \label{eq:classicalIrradiance}
  I_{1,2}(\phi )= I_0 \left[ 1 \pm {\rm Re}\left  (\zeta e^{i\phi}\right ) \right].
\end{equation}
Measurement of $I_{1,2}(\phi )$ at three distinct values of $\phi $
enables independent estimation of the phase and amplitude of
$\Gamma_{12}$. 
When $|\zeta|$ is small, the variance of each measurement of $I_{1,2}$ is approximately equal to the variance of
measurements of $I_0$, given by \cite{ZmuidzinasAO2003}
\begin{equation}
\sigma_I^2= 2 \eta I_0 (1+\eta n) \hbar \omega_c / TA,
\label{ClassicalVariance}
\end{equation}
where $\eta$ is the quantum efficiency of the detector, $n=1/(e^{\hbar \omega_c/k_B\Theta}-1)$ is the mean occupation number at $\omega_c$ for a source with effective temperature $\Theta$, $T$ is the integration time, and $A=\pi W_0^2$ is the cross-sectional area of the optical fields assumed to be in Gaussian modes with beam waist $W_0$. The ``noise equivalent coherence'' corresponding to $|\Gamma_{12,\mathrm{min}}|^2=\sigma_I^2$ is a measure of the minimum detectable mutual coherence in the classical system. $|\Gamma_{12,\mathrm{min}}|$ is independent of $\zeta$, and the signal-to-noise ratio (SNR) in estimating $\Gamma_{12}$ degrades as  $\zeta$ decreases. 

\begin{figure}[t]
        \centering
                \includegraphics[width=8.5cm]{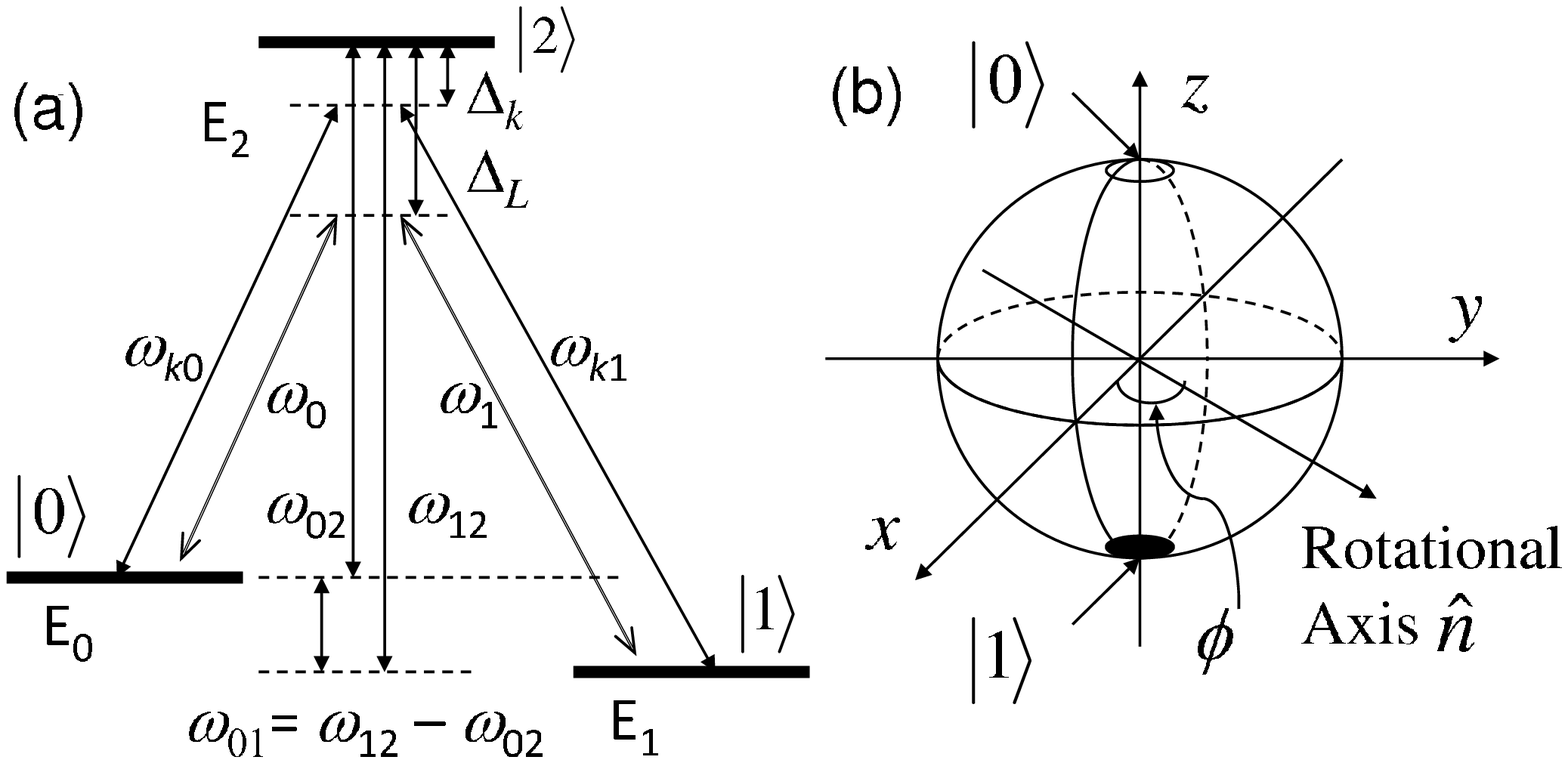}
                \caption{(a) Level structure of the atom under consideration. The two ground states $|0\rangle$ and $|1\rangle$ are coupled via an excited state $|2\rangle$ through optical dipole transitions. $\Delta_k$ and $\Delta_L$  denote the detuning of the field pairs $\hat{c}_{k0}$/$\hat{c}_{k1}$ and  $\hat{d}_{0}$/$\hat{d}_{1}$ from the upper state, respectively. (b) Bloch sphere representation of the atomic system. The interaction with the field rotates the initial state around the axis $\hat{n}$, located on the $x-y$ plane making an angle $\phi$ with the x-axis, by the angle $2|\Omega_0| T$.}
\label{LevelStructure}
\end{figure}

Our model quantum coherence detector replaces the beamsplitter and the two photodetectors (Fig. \ref{InterferometerScheme}b) with a quantum system directly coupled to the two optical fields. As a physical implementation, we consider a collection of three-level atoms in a small optical cavity whose level structure is shown schematically in Fig. \ref{LevelStructure}a. The relevant levels consist of one excited state (denoted by $|2\rangle$) coupled to two ground states ($|0\rangle$ and $|1\rangle$) via Raman transition \cite{RaymerPRA1979}. The transitions between the two ground states and the excited state are chosen to require orthogonal polarization corresponding to the optical fields. We also consider two coherent control beams of frequency $\omega_1$ and $\omega_2$ described by the operators $\hat{d}_1$ and $\hat{d}_2$ interacting with the same atoms, where a complete control of the mutual coherence is available by manipulating the intensities and relative phase of the beams.

The intracavity field $\hat{c}_{k \sigma} \simeq \sqrt{{\cal F}/\pi} \;\; \hat{b}_{k \sigma}$ where ${\cal F}$ is the finesse of the cavity, if the cavity is resonant with the field. The Hamiltonian of the system $\hat{H}  =  \hat{H}_0+\hat{H}_{\mathrm{I}}+\hat{H}_{\mathrm{II}}$ in interaction picture is given by
\begin{eqnarray}
\hat{H}_0 & = &  \sum_{j=0}^2 E_j |j\rangle \langle j|  +  \sum_{k,\sigma}  \hbar \omega_{k \sigma} \hat{c}_{k \sigma}^\dagger \hat{c}_{k\sigma} + \sum_{i=0}^{1}  \hbar \omega_i \hat{d}_i^\dagger \hat{d}_i ,\nonumber  \\
\hat{H}_{\mathrm{I}} & = & \hbar \sum_{k} \left[ g_{k,02}|2\rangle \langle 0| [\hat{c}_{k0} e^{-i\omega_{k0}t} + \hat{c}_{k0}^\dagger e^{i \omega_{k0}t} ] e^{i\omega_{02}t} \right. \nonumber \\
&  +  & \left. \!\! \! g_{k,12}|2\rangle \langle 1| [\hat{c}_{k1}  e^{-i(\omega_{k1}t \!-\! \phi_{k})} \!+\! \hat{c}_{k1}^\dagger  e^{i(\omega_{k1}t \!- \! \phi_k)} ] e^{i\omega_{12}t} \right] \nonumber \\
& + & \!\! h. c., \\
\hat{H}_{\mathrm{II}} & = & \hbar \sum_{i=0}^{1} g_{i2}|2\rangle \langle i| [\hat{d}_i e^{-i\omega_{i}t} +\hat{d}_i^\dagger e^{i\omega_{i}t}] e^{i\omega_{i2}t} + h.c., \nonumber
\label{SystemHamiltonian}
\end{eqnarray}
where $E_j$ is the energy of the atomic state $|j\rangle$, $\omega_{ij} = (E_j-E_i)/\hbar$ is the frequency difference between two atomic states, $\sigma=0$ ($\sigma=1$) denote polarization that induces transition between states $|0\rangle$ ($|1\rangle$) and $|2\rangle$, $\omega_{k\sigma}$ is the energy of the optical mode with momentum $k$ and polarization $\sigma$,   $g_{k,02}$ ($g_{k,12}$) is the coupling coefficient of optical transition between $|0\rangle$ ($|1\rangle$) and $|2\rangle$ via $\hat{c}_{k0}$ ($\hat{c}_{k1}$), $\phi_{k}$ is the phase difference between the two optical fields, and $h.c.$ denotes Hermitian conjugate. The vacuum energy for the photons were suppressed for simplicity.

When $\hat{H}_\mathrm{I} = 0$, this Hamiltonian describes an atomic system interacting with two coherent fields through Raman transition \cite{WinelandJRNIST1998}. For the purpose of utilizing the system as a coherence detector, we extend the analysis to the case when it is driven by partically coherent fields. We consider the evolution of the state described by
\begin{equation}
|\Phi_I (t)\rangle = C_{0}(t)|0\rangle + C_{1}(t)e^{i\omega_{01}t} |1\rangle + C_{2}(t)e^{i\omega_{02}t} |2\rangle.
\label{DynamicState}
\end{equation}
The Schr\"{o}dinger equation $i \hbar \frac{\partial}{\partial t} |\Phi_I (t) \rangle = \hat{H}_\mathrm{I} |\Phi_I (t) \rangle$ is reduced to three differential equations for the coefficients $C_i (t)$. The equation for $C_{2} (t)$ can be formally integrated, and utilized to obtain integro-differential equations for $C_{0} (t)$ and $C_{1} (t)$. Making rotating wave approximation in the limit of large detuning $2 \pi \Delta_k \equiv \omega_{02}-\omega_{k0} (= \omega_{12}-\omega_{k1}) \gg \Delta \omega$ and considering time evolution over timescales much longer than the optical bandwidth ($t \gg 2 \pi / \Delta \omega$)~\cite{CohenTannoudji1977}, the resulting equations reduce to
\begin{eqnarray}
\dot{C}_{0}(t) & \simeq & i \delta_0 C_{0}(t) + i \Omega (t) C_{1}(t) , \nonumber \\
\dot{C}_{1}(t) & \simeq & i \Omega^* (t) C_{0}(t) + i \delta_1 C_{1}(t),
\label{SchrodingerEq2}
\end{eqnarray}
where $\delta_i \simeq \sum_{k k'}g^*_{k,i2} g_{k',i2} \hat{c}_{ki}^\dagger \hat{c}_{k'i}/ (\omega_{i2}-\omega_{k'i})$ are the AC Stark shifts and
\begin{equation}
\Omega (t) \equiv \sum_{kk'} g_{k,02}^* g_{k',12} \frac{\hat{c}_{k0}^\dagger \hat{c}_{k'1} e^{i [(\omega_{k0}-\omega_{k'1})t+\phi_{k'}]}}{\omega_{02}-\omega_{k'1}}
\label{RabiFreq}
\end{equation}
is the (complex) Rabi frequency of the system. Under the large detuning assumption, the Rabi frequency ensemble-averaged over the bandwidth of the optical fields becomes time-independent and reduces to
\begin{eqnarray}
\Omega_0  \equiv \langle \Omega (t) \rangle  & =& \sum_{kk'} g_{k,02}^* g_{k',12} \frac{ \langle\hat{c}_{k0}^\dagger \hat{c}_{k'1}\rangle e^{i [(\omega_{k0}-\omega_{k'1})t+\phi_{k'}]} }{ \omega_{02}-\omega_{k'1}}  \nonumber \\
& \simeq & \frac{Z_0 d_{02} d_{12}}{\hbar^2 \Delta} \frac{\omega_{02} \omega_{12}}{\bar{\omega}_0 \bar{\omega}_1} \frac{{\cal F}}{\pi}\Gamma_{12}(\tau ),
\label{RabiFrequencyEnsemble}
\end{eqnarray}
where $d_{i2}$ denote the dipole moment between the two atomic states $|i\rangle$ and $|2\rangle$, $\Delta \equiv \langle \Delta_k \rangle$ and $\bar{\omega}_i\equiv \langle \omega_{ki} \rangle$ are the ensemble average values of the detuning and the optical frequency, respectively, taken over the optical field, and the phase difference is taken to be $\phi_{k} = \omega_{k1} \tau$. In the second equality, we expressed the electric field using the field annihilation operators \cite{WallsMilburn1994} and used the definition for $\Gamma_{12}(\tau)$. We see that the ensemble-averaged Rabi frequency becomes time-independent since $\langle \hat{c}_{k0}^\dagger \hat{c}_{k'1} \rangle \propto n_k \delta_{kk'}$, and is proportional to the mutual coherence $\Gamma_{12}(\tau)$ of the two optical fields. 
If the atoms are prepared in the initial state 
\begin{equation}
|\Psi (t=0) \rangle = \cos \alpha |0\rangle + e^{i \beta} \sin \alpha |1\rangle,
\label{InitialState}
\end{equation}
one can perform the coordinate transformation $C'_i(t) \equiv C_i (t) e^{-i\delta_i t}$, to find an exact solution to Eqs. (\ref{SchrodingerEq2}).
When the two levels $|0\rangle$ and $|1\rangle$ are exactly degenerate in the presence of AC Stark shifts ($ \delta_0 = \delta_1$),  the system undergoes a coherent Rabi oscillation between the two states. If we express the Rabi frequency as $\Omega_0 \equiv |\Omega_0|e^{i\phi}$, the solution simplifies to
\begin{eqnarray}
\langle C'_0(t) \rangle & = & \cos \alpha \cos |\Omega_0| t + i e^{i(\phi + \beta)} \sin \alpha \sin |\Omega_0| t,\;\;\;\;\;\;\;\;  \label{Solution2}
\\
\langle C'_1(t) \rangle & = & e^{i \beta} \sin \alpha \cos |\Omega_0| t + i e^{-i\phi} \cos \alpha \sin |\Omega_0| t. \nonumber  
\end{eqnarray}

The dynamics of this two-level atomic system can be described as a qubit on a Bloch sphere, where the two states are represented by the south and north pole, respectively (Fig. \ref{LevelStructure}b) \cite{NielsenChuang2000}. The initial state described by Eq. (\ref{InitialState}) is a spin coherent state on the Bloch sphere characterized by the polar angle $2\alpha$ and azimuthal angle $\beta$. The evolution of the atomic state interacting with the optical field over time $T$ corresponds to a rotation around an axis $\hat{n}$ in the $x-y$ plane by the amount $2|\Omega_0| T$. The rotational axis $\hat{n}$ passes through the origin and makes an angle $\phi$ with the $x-$axis. 

The solution given by Eq. (\ref{Solution2}) is formally identical to the case when the system is driven by two coherent states, but with two critical differences. First, the Rabi frequency and  AC Stark shifts are given by the ensemble-averaged value over the partially coherent field modes whereas such averaging is not necessary for  pure coherent states. Second, the variance of the Rabi frequency is non-zero and the atomic system evolution  features fluctuations whereas such fluctuation can be suppressed for the coherent state-driven case. This fluctuation contributes to the SNR of the detection scheme.

Direct measurement of the atomic population after the interaction provides a measure of the observable $|\Gamma_{12}(\tau)|^2$ rather than the mutual coherence $\Gamma_{12}(\tau)$. In order to measure $\Gamma_{12}(\tau)$, we take the approach of adding the two coherent-state control beams to the system to counteract the state evolution caused by the two optical fields under investigation (Fig. \ref{InterferometerScheme}b). The system evolution is described by including all the terms in the Hamiltonian (Eq. \ref{SystemHamiltonian}). The resulting solution is identical in form as Eq. (\ref{Solution2}) with the AC Stark shift terms and the Rabi frequency replaced by
\begin{eqnarray}
\delta_i & \rightarrow & \delta_i' = \delta_i+\delta_i'', \;\; \Omega_0 \; \rightarrow \; \Omega_T = \Omega_0+\Omega_1, \nonumber \\
\delta_i'' & \equiv & |g_{i2}|^2 |\alpha_i|^2 / \Delta_L, \\
\Omega_1 & \equiv & g_{02}^* g_{12} \alpha_0^* \alpha_1 e^{-i \psi}  / \Delta_L, \nonumber
\label{FullSolution}
\end{eqnarray}
where $\alpha_0$ and $\alpha_1$ are the coherent amplitudes of the two coherent fields and $\psi$ is the phase difference between them. In this approach, the atomic system prepared in a set of well-defined states ({\it e.g.}, a few states in the equator of the Bloch sphere) interacts with the two unknown fields and the two control fields for a prescribed time $T$, and then the atomic population is measured in the two basis states $|0\rangle$ and $|1\rangle$ with close to unity quantum efficiency \cite{NagourneyPRL1986}. The amplitudes and relative phase difference between the two coherent control fields are adjusted as the experiment is repeated, until the atomic system evolution is completely suppressed ($\Omega_T=0$). Under this condition, the coherence of the control fields infer the mutual coherence to be measured ($\Omega_0 =-\Omega_1$).

The sensitivity of this approach to coherence detection is limited by four major noise processes: the fluctuation in the Rabi frequency due to the fluctuation of the optical fields  (``photon noise''), the noise associated with determining the coherence of the control beams (``optical readout noise''), the spontaneous emission of the atoms during the Raman process (``spontaneous noise''), and the statistical noise from detection of atomic states, which will be projected to one of the two basis states (``atom shot noise''). The photon noise in atomic evolution is contributed only by the partially coherent fields (and not by coherent control fields), and the variance averaged over the measurement time $T$ is calculated by considering \cite{ZmuidzinasAO2003}
\begin{equation}
\langle \Omega^2 \rangle \! = \! \frac{1}{T^2} \! \int_0^T \!\!\! dt_1 \!\! \int_0^T \!\!\! dt_2 \langle \Omega(t_1) \Omega(t_2) \rangle \nonumber \! \simeq \! \Omega_0^2 (1 \!+ \!\frac{2 \pi}{T \Delta \omega} \frac{n \!+ \!1}{n}). 
\label{variance}
\end{equation}
Based on Eq. (\ref{RabiFrequencyEnsemble}), the variance in measurement of the mutual coherence due to photon noise is given by $\sigma_p^2 = |\Gamma_{12}|^2 (2 \pi / T \Delta \omega)(n+1)/n$. The shot noise corresponding to determining the mutual coherence of the control beams is given by $2 I_L (\hbar \omega_L/TA)$, where $I_L$ is the intensity of the laser beam interacting with the atoms. Under the balance condition $I_L/\Delta_L \sim  \Gamma_{12}{\cal F}/\pi \Delta$, the optical readout noise for the mutual coherence is given by
\begin{equation}
\sigma_L^2= 2 I_L \frac{\hbar \omega_L}{TA} \left(\frac{\Delta}{\Delta_L}\frac{\pi}{{\cal F}}\right)^2  \simeq 2 |\zeta| n \left( \frac{\hbar \omega_c}{TA}\frac{\pi}{{\cal F}}\right)^2 \frac{\Delta \omega T}{2 \pi} \frac{\Delta}{\Delta_L} ,\nonumber
\label{OpticalReadoutNoise}
\end{equation} 
where we assume $\omega_{1,2} \simeq \omega_c \simeq \omega_L$. For the spontaneous noise, we require that the spontaneous emission rate of the upper state of the atom $\gamma = d_{02(12)}^2\omega_{02(12)}^3/3 \pi \epsilon_0 \hbar c_0^3$ ($\epsilon_0$ is the electrical permittivity and $c_0$ is the speed of light in vacuum) be small compared to the Rabi frequency due to the optical fields, so that the ``error'' from spontaneous emission during the coherence measurement is small. The corresponding noise is given by $\sigma_{sp} \simeq (4\pi^2 \hbar \omega_c \Delta / 3 \lambda^2 {\cal F})^2$, where $\lambda=2 \pi / \omega_c$ is the central wavelength of the optical fields. For the atom shot noise, the precision with which one can determine the angle of system rotation is given by the shot noise in the number of atoms detected in each state. The corresponding variance in mutual coherence is given by $\sigma_a^2 \simeq (\hbar^2 \Delta / d_{02} d_{12}Z_0T)^2 /N_a$, where $d_{i2}$ is the optical transition dipole moment between the state $|i\rangle$ and $|2\rangle$. The overall variance of mutual coherence detection in the quantum coherence detector is given by $\sigma_Q^2 = \sigma_p^2+\sigma_L^2+\sigma_{sp}^2 + \sigma_a^2$.

\begin{figure}[t]
        \centering
                \includegraphics[width=8cm]{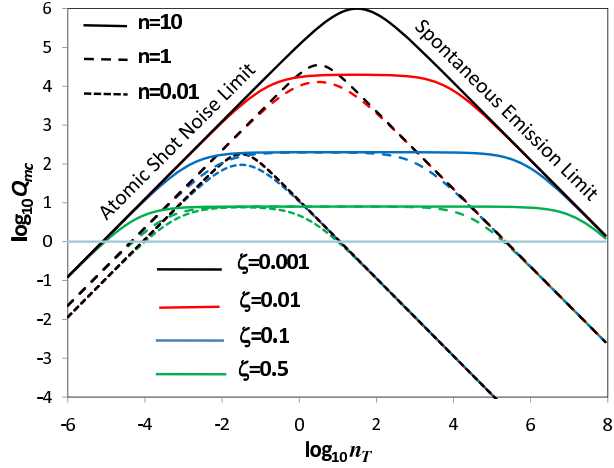}
                \caption{Enhancement factor $Q_{mc}$ of quantum coherence sensor over classical interference detectors plotted as a function of $n_T$. Solid, dashed and dotted lines represent the average number of photons per mode  $n=10$, $1$ and $0.01$, respectively, while the green, blue, red and black curves correspond to $\zeta$ values of 1, 0.1, 0.01 and 0.001, respectively.}
\label{SensitivityPlots}
\end{figure}

In the Gaussian beam geometry, the total number of atoms in the beam waist $W_0$ within Rayleigh length is given by $N_a \sim 2 N_0 \pi^2 W_0^4/\lambda = 2N_0 A^2/\lambda$, where $N_0$ is the density of atoms. We define the sensitivity enhancement factor $Q_{mc}\equiv \sigma_I^2/\sigma_Q^2$ over classical interferometric approach as
\begin{eqnarray}
Q_{mc} & \simeq & \!\! \left[ \frac{ |\zeta|^2}{2} \!+\! \left(\frac{\pi}{{\cal F}}\right)^2 \left\{\frac{|\zeta|\Delta}{2(n\! + \!1)\Delta_L} \!+\! \frac{32 \pi^4 \Delta^2 A^2 n_T}{9 \lambda^4 n^2 (n\!+\!1) \Delta \omega^2} \right. \right. \nonumber \\
&& \left. \left. + \left(\frac{\hbar}{Z_0 d_{02} d_{12}}\frac{\Delta}{\omega_c}\right)^2 \frac{\lambda}{4 N_0 (1 + n)n_T } \right\} \right]^{-1}, 
\label{Comparison}
\end{eqnarray}
where $n_T = n \Delta \omega T/ 2 \pi$ is the total number of photons detected per measurement time interval $T$, and an efficient irradiance detector was assumed ($\eta=1$). Figure \ref{SensitivityPlots} shows $Q_{mc}$ as a function of $n_T$ assuming small volume cavity with high finesse, using typical values of $d_{02} = d_{12} \simeq 1.6 \times 10^{-29}$Cm,  $\Delta \omega \sim 10^{-7}\omega_c$, $\lambda = 780$nm, $N_0 \sim 10^{18} \mathrm{m}^{-3}$, $W_0=1 \; \mu$m, and {\cal F} = 1.6$\times 10^5$.

At low $n_T$, the sensitivity of the quantum coherence detector is limited by the atom shot noise, and is below that of classical interferometric detector. As the  average photon number per mode $n$ increases, the thermal noise in the optical fields increase and the quantum coherence detector performance improves as the SNR in classical measurement degrades (dotted, dashed and solid curves). The noise performance of the quantum detector improves as $n_T$ increases, until it is limited by the optical readout noise or the photonic noise limit. In this limit, the dependence of noise on $n_T$ is identical for the quantum and classical case, so the enhancement factor $Q_{mc}$ remains constant as a function of $n_T$. In classical interferometry, the noise is determined by the photon noise corresponding to the irradiance of each optical field ($\Gamma_{11}$ and $\Gamma_{22}$), while in the quantum case the noise contribution only comes from the amount of mutual coherence present between the two fields. When the mutual coherence is very small between two fields with low average photon number per mode, the quantum coherence detector sensitivity is higher by a factor of $1/\zeta^2$. At higher values of $n_T$, longer integration time increases the probability of spontaneous emission and the senstivity is determined by the spontaneous emission noise.

In this paper, we showed that the mutual coherence of two optical fields can be directly measured using an atomic system when quantum-mechanically coherent states can be prepared, with performance exceeding classical detection schemes. The authors would like to thank Daniel Gauthier, Stephen Teitsworth and Bernard Yurke for helpful discussions. This program was supported by DARPA through AFOSR Grant \#FA9550-06-1-0230.


\end{document}